\PassOptionsToPackage{numbers,sort&compress}{natbib}
\documentclass{article}

\usepackage{graphicx}

\usepackage[creativeai, final]{neurips_2025}

\usepackage{lineno}
\usepackage{float}
\usepackage{natbib}
\usepackage{geometry}
\usepackage{amsmath}
\usepackage[utf8]{inputenc} 
\usepackage[T1]{fontenc}    
\usepackage{hyperref}       
\usepackage{url}            
\usepackage{booktabs}       
\usepackage{amsfonts}       
\usepackage{nicefrac}       
\usepackage{microtype}      
\usepackage{xcolor}         
\usepackage{flafter}
\usepackage{placeins}
\usepackage{comment}
\setcitestyle{numbers,square}

\title{LUMIA: A Handheld Vision-to-Music System for Real-Time, Embodied Composition}

\author{}
\usepackage{authblk}

\author[1]{Connie Cheng\textsuperscript{*}}
\author[1]{Chung-Ta Huang\textsuperscript{*}}
\author[2]{Vealy Lai}
\affil[1]{Harvard University, Cambridge, MA 02138\\
\texttt{\mbox{connie\_cheng@gsd.harvard.edu, chungta\_huang@gsd.harvard.edu}}}
\vspace{1.5em}
\affil[2]{Massachusetts Institute of Technology, Cambridge, MA 02139\\
\texttt{laiv@mit.edu}}

\begin{document}

\maketitle

\begin{abstract}
Most digital music tools emphasize precision and control, but often lack support for tactile, improvisational workflows grounded in environmental interaction. Lumia addresses this by enabling users to "compose through looking"—transforming visual scenes into musical phrases using a handheld, camera-based interface and large multimodal models. A vision-language model (GPT-4V) analyzes captured imagery to generate structured prompts, which, combined with user-selected instrumentation, guide a text-to-music pipeline (Stable Audio). This real-time process allows users to frame, capture, and layer audio interactively, producing loopable musical segments through embodied interaction. The system supports a co-creative workflow where human intent and model inference shape the musical outcome. By embedding generative AI within a physical device, Lumia bridges perception and composition, introducing a new modality for creative exploration that merges vision, language, and sound. It repositions generative music not as a task of parameter tuning, but as an improvisational practice driven by contextual, sensory engagement~\ref{fig:device}. \footnotetext{Project page \& code:
\url{https://github.com/KidaGSD/LLOv2}}

\end{abstract}

\begin{figure}
    \centering
    \includegraphics[width=0.6\linewidth]{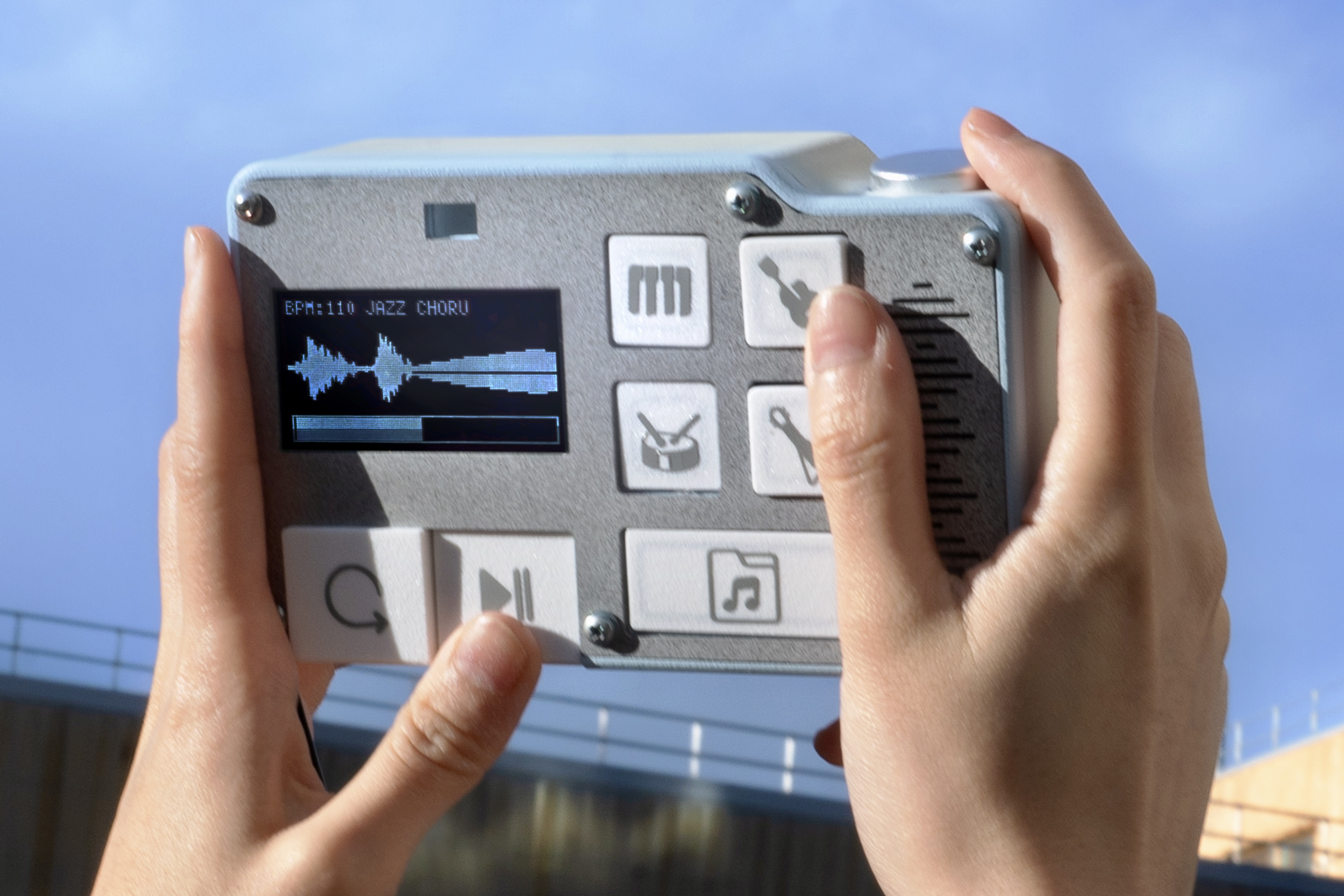}
    \caption{Lumia Device}
    \label{fig:device}
\end{figure}

\section{Introduction}
Recent advances in generative AI have enabled creative tools across text, image, and audio, yet most remain screen-based and prompt-driven, limiting physical engagement and real-time improvisation in music production. \textit{Lumia} is a handheld, camera-based device for real-time music generation from visual input. It uses GPT-4 Vision to analyze captured scenes and construct structured prompts from objects, context, and inferred mood, which are then passed to Stable Audio to synthesize short loops~\citep{openai2024gpt4technicalreport,StableAudio2}. Users influence generation by framing images, selecting instruments, and layering clips via a browser or physical controller, while the model introduces interpretive variation. \textit{Lumia} extends the Large Language Object (LLO) concept~\citep{Coelho2024LLO}, as introduced in VBox~\citep{Liang2024VBox}, which embedded generative models in materially expressive systems. Whereas VBox enabled tactile navigation of latent audio spaces, \textit{Lumia} shifts to composition, linking visual perception with multimodal generation. It treats the visual world as a source of sonic material, enabling a co-creative workflow where user intent and model inference converge, resulting in an embodied, improvisational interface for AI-driven sampling and composition grounded in environmental context.

\section{Related Work}
\subsection{Multimodal Generative Models}
Recent models such as DALL·E, and GPT-4 Vision have shown strong performance in cross-modal tasks, including image captioning, scene understanding, and prompt generation~\citep{ramesh2022hierarchical}. In the audio domain, models like MusicLM, AudioLDM, and Stability AI’s Stable Audio have demonstrated the feasibility of text-to-music generation with varying degrees of control over structure, instrumentation, and genre~\citep{agostinelli2023musiclmgeneratingmusictext,audioldm2023,StableAudio2}. While these systems demonstrate strong generative performance, they are typically accessed via prompt-based or batch-processing interfaces, and offer limited support for interactive scenarios such as live composition, exploratory sampling, or iterative refinement. Most lack mechanisms for real-time control over generation parameters, continuous multimodal input~\citep{AudioCLIP2021}, or feedback-driven adaptation, which are critical for workflows that depend on responsiveness, improvisation, and embodied engagement.
\subsection{Creative AI and Co-Creation}
Work in creative AI has increasingly emphasized tools that position the human as an active collaborator rather than a passive consumer. Systems such as Magenta Studio, Jukebox, and Soundify offer musicians new modes of engagement, but often assume technical familiarity or require traditional DAW integration~\citep{Magenta2016,jukebox2020}. Lumia differs by embedding creative interaction into a physical object, making it more accessible and intuitive, especially for non-experts. It aligns with ongoing research into co-creative systems, where human input and machine generation occur in tandem, influencing one another in real time\citep{Amershi2019Guidelines}.
\subsection{Tangible and Embodied Interfaces}
There is a long-standing tradition in HCI of exploring tangible and embodied interfaces for creative expression. Systems such as Reactable and Bela-based musical hardware~\citep{Jorda2007Reactable,McPherson2015Bela} have shown that hands-on interaction can enhance improvisation, flow, and expressive control in music-making. These systems often operate in fixed environments and rely on predefined spatial mappings. More recent work, such as Be the Beat\citep{10.1145/3689050.3705995}, extends this paradigm by embedding generative AI into a beatbox-shaped device that responds to dancers’ movements, using physical form language to cue interaction and support situated, improvisational performance. Lumia builds on this direction by introducing a mobile, camera-inspired interface for visual exploration and sound generation, further emphasizing physical context and environmental engagement as central components of co-creative AI systems.
\subsection{Gap and Positioning}
\begin{figure}
    \centering
    \includegraphics[width=0.8\linewidth]{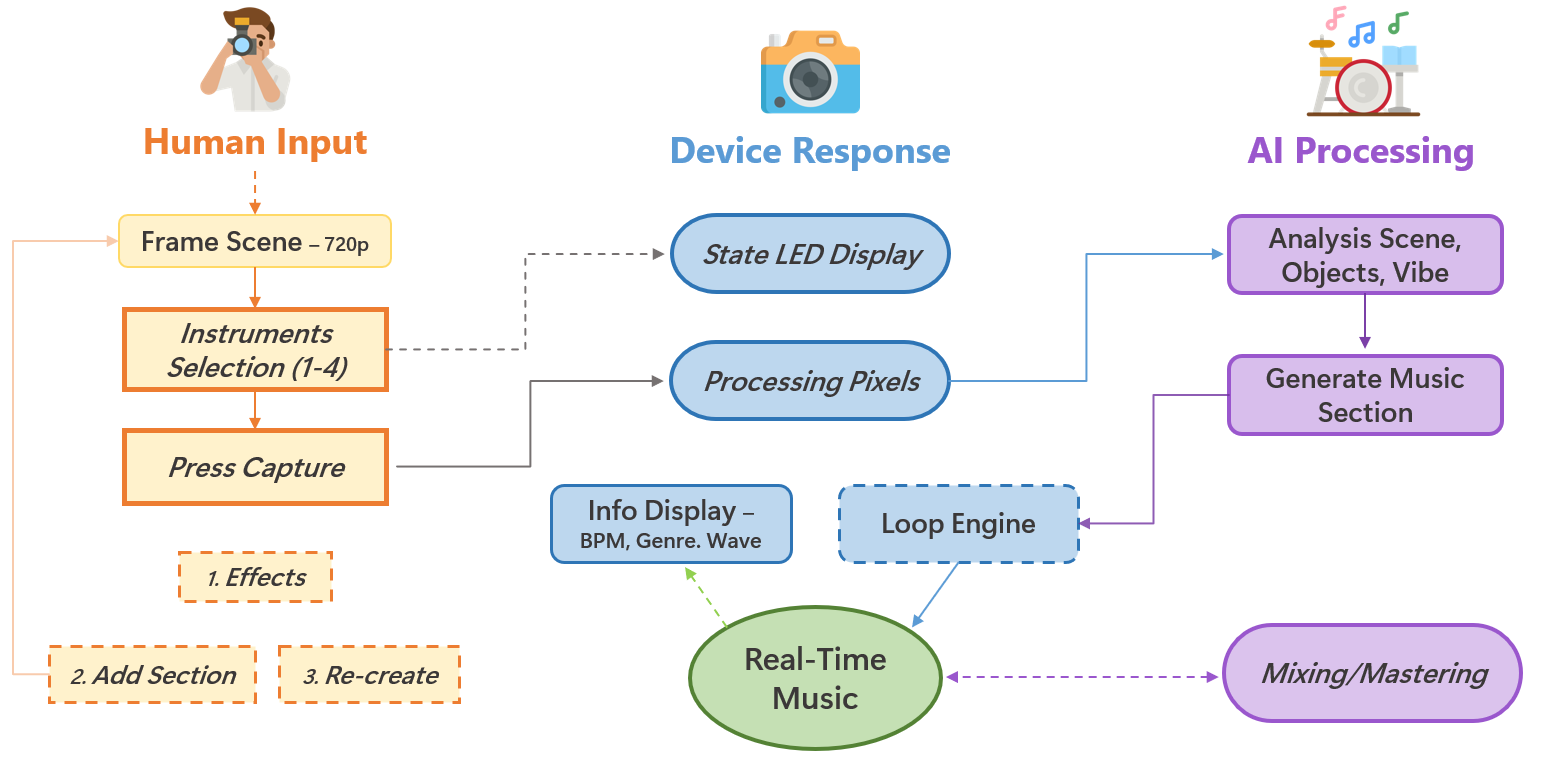}
    \caption{User Flow}
    \label{fig:userflow}
\end{figure}

While prior work has addressed text-to-music generation, multimodal synthesis, and tangible instruments, few systems combine these elements into a real-time, embodied workflow grounded in physical and environmental interaction. Most existing tools remain screen-based or prompt-driven, limiting their use in improvisational or situated creative practices. Lumia fills this gap by framing vision as a form of sampling and embedding generative AI into a portable, physically expressive device. It enables users to compose music through visual exploration, positioning AI as a responsive collaborator in a context-aware, co-creative process(see Figure~\ref{fig:userflow}).

\section{Methodology}
Lumia was designed to enable music creation from visual input while preserving user agency, with large multimodal models forming the core of its visual-to-audio pipeline. To keep interaction intuitive and accessible, it generates fixed-length audio clips that users can blend and sequence into tracks. Inspired by DJ turntables, early prototypes focused on live layering and looping, exploring interaction styles where clips were added sequentially, like a conductor introducing instruments. This led to experiments with single-instrument samples for flexible mixing, while a physical turntable informed timing and blending dynamics. Rapid prototyping and informal testing shaped the camera-inspired form and audio-themed visual language. Iterative development revealed key insights for coherence: image color strongly influenced genre inference, prompting the addition of physical color filters for simple visual genre control; overemphasis on literal objects degraded audio, so prompts were biased toward contextual and atmospheric descriptors, with genre, tempo, and key kept consistent across segments. Technical challenges included harmonizing new samples with prior material—the generation model’s lack of temporal awareness made full-track regeneration disruptive. This was addressed with a sequential composition model, appending each new section in time to preserve responsiveness and continuity.

\subsection{Physical Form and Interaction}
The device resembles a compact camera with five buttons: four for selecting instruments (keys, guitar, bass, percussion) and one for capture and playback. LEDs indicate the current instrument state, and an onboard display shows tempo, genre, section role, and audio levels. A built-in speaker handles real-time loop playback, while a microcontroller manages I/O. A camera module continuously streams video to the frontend; still frames are sent to a vision-language model (VLM) for analysis.

\subsection{Hardware Protocol and Microcontroller Firmware}
The Arduino firmware is implemented as a finite-state loop. It debounces button inputs, updates the display, and adjusts LEDs and audio level indicators. LCD updates reflect session metadata, while LEDs respond to both software state and user input. The firmware is stateless beyond its display responsibilities, making all playback and audio logic frontend-driven.

\section{System Architecture}
\begin{figure}
    \centering
    \includegraphics[width=1\linewidth]{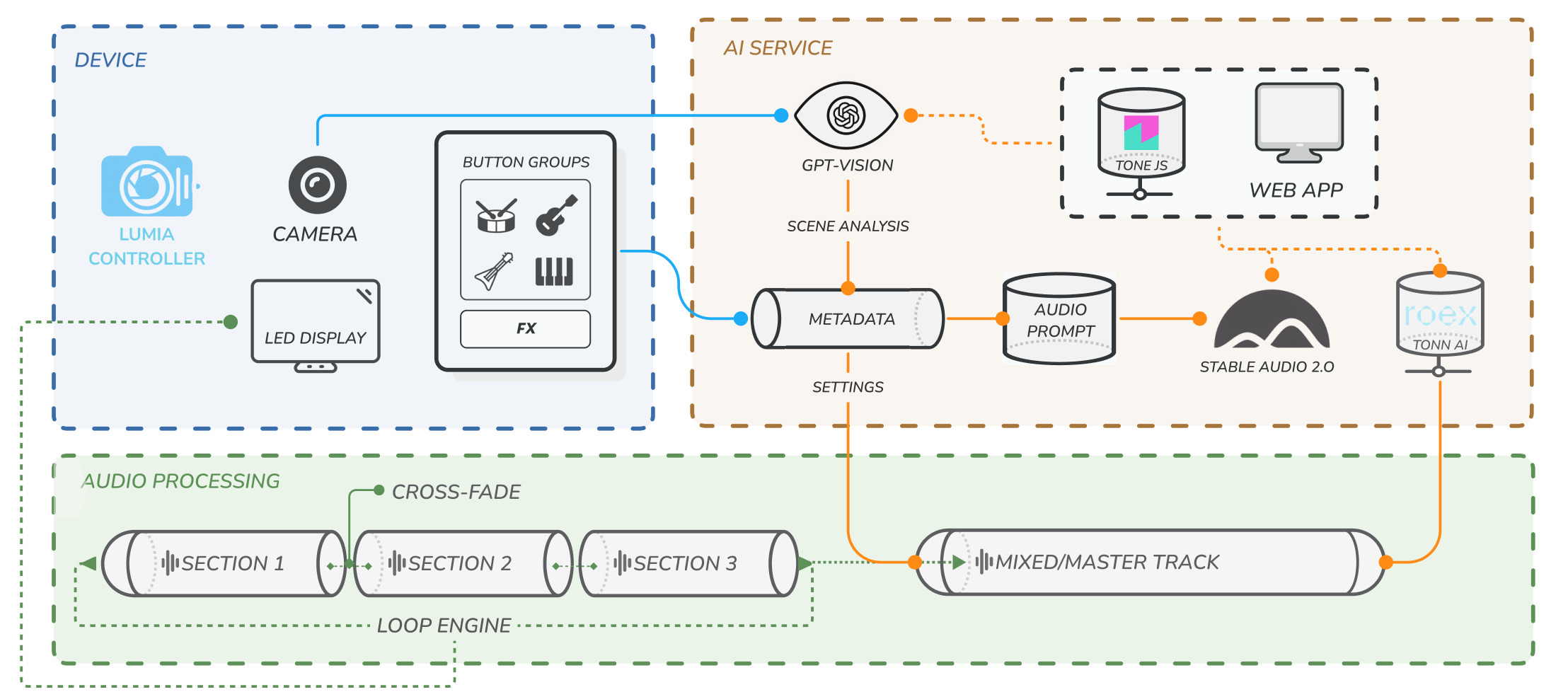}
    \caption{System Diagram}
    \label{fig:system}
\end{figure}

Lumia’s system architecture (see Figure~\ref{fig:system}) is designed as a modular, decoupled system comprising three primary pillars: (1) a tangible hardware controller for physical interaction, (2) a browser-based frontend application that serves as the central orchestrator, and (3) a suite of cloud-based AI services for content generation and processing. This architecture is designed for instantaneous physical feedback to the user while managing the high-latency, asynchronous nature of generative AI calls in the background, ensuring a natural creative experience. Full details of hardware configuration, API service specifications, latency, and cost are provided in Appendix to support reproducibility.

\subsection{Music Generation: Vision-to-Music Pipeline}
Image Captioning: When the capture button is pressed, a frame from the live camera feed is sent to GPT-4 Vision for extracting meaningful scene information. The vision model returns a structured JSON caption describing the image, with fields for: 1) overall description of the scene, 2) a list of salient objects, 3) the overall mood(adjectives), 4) section role $\in\{\text{\emph{intro}, \emph{verse}, \emph{chorus}, \emph{bridge}, \emph{outro}}\}$, 5) a music genre(based on the first section), 6) suggested BPM(beats per minute).  This metadata is designed to capture both the content and the atmosphere of the photo, providing a basis for music generation. (If section inference is uncertain, system would default to \emph{verse})

\subsection{Prompt Construction:}
The system constructs a music generation prompt by combining structured GPT-4 Vision output with user-defined instrumentation. Users select 1 to 3 instruments (keys/synth, guitar, bass, drums) to define the sonic range; this cap promotes perceptual stream segregation and reduces masking in novice-facing settings~\cite{Bregman1990}. The parsed image caption containing scene description, mood, genre, BPM, and section role (e.g., intro, verse, chorus)—is programmatically merged with these inputs. To scaffold musical structure, section-specific modifiers (e.g., “higher energy, catchy hookc for chorus; “winding down” for outro) and, for non-initial segments, a variation tag (e.g., “motif development”, “steady groove”) are appended to promote temporal progression and thematic continuity. All elements are concatenated into a single sentence-level prompt optimized for the Stability AI audio generation model. For example:
\begin{quote}\small\ttfamily
keys, guitar section, purple neon street light sign at night, moody, lush, ambient chill, steady groove, subtle variation, same sound palette as previous section
\end{quote}
This structured prompt encodes intended instrumentation, musical character, temporal role, and stylistic direction to condition the audio model effectively.

\subsection{Audio Generation}
 The composed prompt is sent to the Stability AI-Stable Audio API(2.0) for music generation. Each section is generated as a 15 seconds audio clip with a target tempo derived from the caption. The generation parameters (stereo output at 44.1kHz) are held constant for all sections to ensure uniform audio quality. The Stable Audio model interprets the text prompt to produce an audio waveform matching the described scene, mood, instruments, and genre. The returned audio WAV file is immediately decoded into an AudioBuffer and added to the playback loop.

\subsection{Loop Playback Engine}

\textbf{Timing and scheduling: }
Once the first section is generated, the engine enters a continuous loop.
Let $b$ denote the session tempo (BPM).
A beat lasts $T_{\mathrm{beat}}=60/b$ seconds and a bar is $T_{\mathrm{bar}}=4T_{\mathrm{beat}}$.
We generate fixed-length sections $L_k=m_kT_{\mathrm{bar}}$, $m_k\in\mathbb{N}$.
To ensure smooth overlaps we define a tempo-adaptive crossfade window (Eq.~\ref{eq:tcf})
and schedule the next section start time (Eq.~\ref{eq:sched-loop}):
\noindent
\begin{minipage}[t]{0.55\linewidth}
\begin{equation}
T_{\mathrm{cf}}(b)=\max\!\Bigl(\tfrac{120}{b},\,0.3\Bigr)\ \text{s}
\label{eq:tcf}
\end{equation}
\end{minipage}\hfill
\begin{minipage}[t]{0.42\linewidth}
\begin{equation}
t_{k+1}=t_k + L_k - T_{\mathrm{cf}}
\label{eq:sched-loop}
\end{equation}
\end{minipage}
\textbf{Crossfade envelopes: }
To avoid clicks and maintain perceived loudness during the overlap, we default to equal-power(eq) crossfades.
For outgoing buffer $x[n]$ and incoming buffer $y[n]$ we set
\begin{equation}
g_{\mathrm{out}}(n)=\cos\!\Bigl(\tfrac{\pi n}{2N}\Bigr),\quad
g_{\mathrm{in}}(n)=\sin\!\Bigl(\tfrac{\pi n}{2N}\Bigr),\quad
z[n]=g_{\mathrm{out}}(n)x[n]+g_{\mathrm{in}}(n)y[n],
\label{eq:eqpower}
\end{equation}
Since $g_{\mathrm{out}}^2(n)+g_{\mathrm{in}}^2(n)=1$, the instantaneous power remains approximately constant.
For sparser, ambient sections with small energy changes, we optionally use a single power-law(poly) envelope
\begin{equation}
g_{\mathrm{out}}(n)=\Bigl(1-\tfrac{n}{N}\Bigr)^{\alpha_0},\qquad
g_{\mathrm{in}}(n)=\Bigl(\tfrac{n}{N}\Bigr)^{\alpha_0},
\label{eq:poly-fades}
\end{equation}
with hyperparameter $\alpha_0\approx 2.5$, providing a smoother, less hazy transition. Alternative shapes can be swapped in using the same scheduling logic. 

\textbf{Context-to-envelope policy: }
We map a lightweight context vector $\mathbf{c}=(\Delta P,\,\text{section role})$ to a fade family and parameter.
Conceptually, envelope selection can be framed as minimizing a loudness mismatch objective,
\vspace{-3mm}
\begin{equation}
(f^\star,\theta^\star)
=\arg\min_{f\in\{\text{eq},\,\text{poly}\},\,\theta}
\sum_{n=0}^{N-1}\bigl(|z[n]|^2 - P_{\mathrm{target}}\bigr)^2
+ \lambda\,\mathcal{C}_{\mathrm{transient}},
\label{eq:policy}
\end{equation}
where $P_{\mathrm{target}}$ is the running power target and $\mathcal{C}_{\mathrm{transient}}$ penalizes pre/post‑splice transients.

\textbf{Look‑ahead and hot‑swap: }
A short look‑ahead scheduler quantizes all splice times to the nearest bar boundary.
This preserves the groove and allows “hot‑swaps”: when an asynchronous preview mix or mastered version becomes available (see Section~\ref{sec:mixing}), the engine atomically switches sources at the next bar boundary with uninterrupted, beat-aligned playback.

\section{Automatic AI Mixing and Integration}
\label{sec:mixing}

\paragraph{Mixing:} When at least two sections are ready \emph{and} \textit{auto-mix} is enabled, Lumia triggers an asynchronous preview mix job while playback continues. The backend integrates the Tonn AI~\citep{TonnAPI} for multitrack mixing; all calls are non-blocking and results are inserted at bar boundaries in the Loop Engine. First, section WAVs and the Loop Engine files are uploaded as stems. The system then calls \emph{mixpreview} with per-stem metadata, including \texttt{instrumentGroup}, \texttt{presenceSetting}, \texttt{panPreference}, \texttt{reverbPreference}, and \texttt{musicalStyle} derived from the session genre. A \emph{task ID} is returned, and completion is detected via webhook. Once ready, the preview mix is downloaded, decoded to an \texttt{AudioBuffer}, and scheduled for a hot-swap at the next loop boundary with  crossfade.

\paragraph{Mastering.}
For export-quality output, sections are first concatenated into a single stereo WAV using \textit{pydub}'s \textit{"AudioSegment"}. This file is then submitted to Tonn AI’s album-style mastering service using the \emph{mastering preview} request, which specifies the musical style, desired loudness, and sample rate (44.1\,kHz). The system retrieves the resulting preview master, and the final master version. The mastered audio is decoded and, like preview mixes, is inserted into the loop at a bar boundary. If additional sections are added while a mastered preview is playing, the system appends the new material and re-runs the mastering process asynchronously without interrupting playback~\citep{TonnAPI,PyDub}.

\section{Experiments}
We conducted a formative evaluation with three professional audio engineers (4--6 years experience). Each participant used \textit{Lumia} to compose a 120--150\,s multi-section track by framing scenes, selecting instruments, and layering loops on-device. Sessions lasted $\sim$25--30\,min and concluded with a short survey and semi-structured interview. The survey assessed five constructs: \emph{co-creation/agency}, \emph{musical quality}, \emph{audio mapping}, \emph{interaction/flow}, and \emph{value/fit}, measured with multi-item Likert-type scales (1--7). Two additional 0--10 sliders measured \emph{authorship share} and \emph{expectation match}. Construct scores were computed as item means, following standard practice for summated ratings~\citep{Likert1932,Carifio2008Likert}. The instrumented workflow and several survey items were adapted from Creative-AI studies such as PAGURI, a user-experience study of creative interaction with text-to-music models~\citep{PAGURI2024}, and modified for \textit{Lumia}'s real-time photo-to-music loop.  

Mean construct scores are provided in (Figure 10: Collected numerical evaluation results\ref{fig:evaluation}). On the 0--10 scales, \emph{authorship} averaged 4.0 and \emph{expectation match} averaged 6.3. Participants cited speed (``Is different started from images rather than a DAW template, this gives faster vibe-finding'') and nuanced control (``Chorus lift with subtle bass change'', ``macro close-up calmed the drums''). Improvement requests included: (i) optional genre/BPM locks, (ii) micro-nudge layer edits, (iii) a visual-to-sonic mapping legend, and (iv) reduced latency. Reported use-cases ranged from rapid idea sketching to mood-boarding and creating backing tracks for short-form video.

\section{Discussion and Future Work}
Lumia demonstrates that multimodal language models, traditionally used for captioning or classification, can operate in real-time generative contexts, translating vision into structured musical prompts while preserving user intent. Informal use suggests it behaves more like an instrument than a static tool, supporting exploration, intuition, and creative ownership. Its camera-inspired physical design plays a critical role in shaping interaction, encouraging playful experimentation and environmental engagement. Users valued the balance of control and surprise in composing through looking, noting patterns in visual-to-audio mappings. Although Lumia offers a real-time generative workflow, several limitations remain. Planned hardware updates include extending the FX button for real-time modulation of pitch, duration, reverb, and volume, mapped to gestures such as shakes, strikes, or pressure input. The current reliance on open-ended scene analysis introduces variability and interpretability challenges; future versions may add intermediate prompt-editing layers to let users refine or constrain descriptions before generation.

Lumia’s accessibility positions it as a tool for inclusive, co-creative music-making, education, and rapid prototyping, expanding opportunities for interdisciplinary practice. Embedding generative AI in cultural production carries risks such as overreliance on machine output, erosion of local styles through model bias, and dependence on proprietary services. These risks can be mitigated through transparency in how Lumia’s visual-to-audio mappings and prompts are constructed, robust user control over generative parameters, and active community participation in shaping system evolution to ensure representation, cultural diversity, and genuine human–AI co-creation.

\section{Conclusion}
We presented Lumia, a camera-inspired handheld system for real-time, vision-driven musical composition. Its core contribution is a real-time vision-to-audio pipeline that uses GPT-4 Vision to transform captured imagery into structured musical prompts, rendered into coherent audio segments by Stability AI’s text-to-music model. This is embedded in a camera-inspired physical interface that uniquely integrates scene framing, instrument selection, and audio layering into a continuous creative loop, enabling responsive, context-aware composition outside of screen-based workflows. The system’s design philosophy prioritizes human–AI co-creation, supporting improvisation, expressive control, and cultural diversity while leveraging the generative capabilities of multimodal models. Together, these elements establish Lumia as the first portable, vision-driven music generation device to merge multimodal AI inference, tangible interaction, and live compositional practice into a unified, user-centered workflow.

\section{Acknowledgements}
We would like to thank Prof. Marcelo Coelho for his guidance and support throughout the development of this project. We are also grateful to our teaching assistants Sergio Mutis, Chenyue Dai, and Quincy Kuang for their feedback and technical insight as well. Special thanks to William McKenna for his mentorship and help with tools around the shop, and to the MIT MAD program and the 4.043/4.044 Design Studio spaces, where the 4.044 community of peers and instructors fostered a wonderful space for experimentation, iteration, and creative risk-taking.

\bibliographystyle{abbrvnat}
\bibliography{lumia_references}

@article{ramesh2022hierarchical,
      title={Hierarchical Text-Conditional Image Generation with CLIP Latents}, 
      author={Aditya Ramesh and Prafulla Dhariwal and Alex Nichol and Casey Chu and Mark Chen},
      year={2022},
      eprint={2204.06125},
      archivePrefix={arXiv},
      primaryClass={cs.CV},
      url={https://arxiv.org/abs/2204.06125}, 
}

@misc{agostinelli2023musiclmgeneratingmusictext,
      title={MusicLM: Generating Music From Text}, 
      author={Andrea Agostinelli and Timo I. Denk and Zalán Borsos and Jesse Engel and Mauro Verzetti and Antoine Caillon and Qingqing Huang and Aren Jansen and Adam Roberts and Marco TagliasFacchi and Matt Sharifi and Neil Zeghidour and Christian Frank},
      year={2023},
      eprint={2301.11325},
      archivePrefix={arXiv},
      primaryClass={cs.SD},
      url={https://arxiv.org/abs/2301.11325}, 
}

@inproceedings{audioldm2023,
      title={AudioLDM: Text-to-Audio Generation with Latent Diffusion Models}, 
      author={Haohe Liu and Zehua Chen and Yi Yuan and Xinhao Mei and Xubo Liu and Danilo Mandic and Wenwu Wang and Mark D. Plumbley},
      year={2023},
      eprint={2301.12503},
      archivePrefix={arXiv},
      primaryClass={cs.SD},
      url={https://arxiv.org/abs/2301.12503}, 
}

@misc{StableAudio2,
      title={Stable Audio Open}, 
      author={Zach Evans and Julian D. Parker and CJ Carr and Zack Zukowski and Josiah Taylor and Jordi Pons},
      year={2024},
      eprint={2407.14358},
      archivePrefix={arXiv},
      primaryClass={cs.SD},
      url={https://arxiv.org/abs/2407.14358}, 
}

@misc{AudioCLIP2021,
      title={AudioCLIP: Extending CLIP to Image, Text and Audio}, 
      author={Andrey Guzhov and Federico Raue and Jörn Hees and Andreas Dengel},
      year={2021},
      eprint={2106.13043},
      archivePrefix={arXiv},
      primaryClass={cs.SD},
      url={https://arxiv.org/abs/2106.13043}, 
}

@article{Coelho2024LLO,
author = {Coelho, Marcelo and Labrune, Jean-Baptiste},
title = {Large Language Objects: The Design of Physical AI and Generative Experiences},
year = {2024},
issue_date = {July - August 2024},
publisher = {Association for Computing Machinery},
address = {New York, NY, USA},
volume = {31},
number = {4},
issn = {1072-5520},
url = {https://doi.org/10.1145/3672534},
doi = {10.1145/3672534},
journal = {Interactions},
month = jun,
pages = {43–48},
numpages = {6}
}

@misc{Liang2024VBox,
  title        = {VBox: AI-Powered Radio for Musical Exploration},
  author       = {Liang, Danning and Laptiev, Artem and Coelho, Marcelo},
  howpublished = {\url{https://doi.org/10.1145/3672534}}, 
  year         = {2024},
  note         = {Referenced in LLO feature}
}

@inproceedings{Jorda2007Reactable,
author = {Jord\`{a}, Sergi},
title = {The reactable: tangible and tabletop music performance},
year = {2010},
isbn = {9781605589305},
publisher = {Association for Computing Machinery},
address = {New York, NY, USA},
url = {https://doi.org/10.1145/1753846.1753903},
doi = {10.1145/1753846.1753903},
booktitle = {CHI '10 Extended Abstracts on Human Factors in Computing Systems},
pages = {2989–2994},
numpages = {6},
location = {Atlanta, Georgia, USA},
series = {CHI EA '10}
}

@inproceedings{Amershi2019Guidelines,
author = {Amershi, Saleema and Weld, Dan and Vorvoreanu, Mihaela and Fourney, Adam and Nushi, Besmira and Collisson, Penny and Suh, Jina and Iqbal, Shamsi and Bennett, Paul N. and Inkpen, Kori and Teevan, Jaime and Kikin-Gil, Ruth and Horvitz, Eric},
title = {Guidelines for Human-AI Interaction},
year = {2019},
isbn = {9781450359702},
publisher = {Association for Computing Machinery},
address = {New York, NY, USA},
url = {https://doi.org/10.1145/3290605.3300233},
doi = {10.1145/3290605.3300233},
booktitle = {Proceedings of the 2019 CHI Conference on Human Factors in Computing Systems},
pages = {1–13},
numpages = {13},
location = {Glasgow, Scotland Uk},
series = {CHI '19}
}

@article{openai2024gpt4technicalreport,
  author       = {OpenAI and Achiam, Josh and others},
  title        = {GPT-4 Technical Report},
  journal      = {arXiv preprint arXiv:2303.08774},
  year         = {2024},
  url          = {https://arxiv.org/abs/2303.08774}
}

@book{Bregman1990,
  author    = {Bregman, Albert S.},
  title     = {Auditory Scene Analysis: The Perceptual Organization of Sound},
  year      = {1990},
  publisher = {MIT Press}
}

@article{PAGURI2024,
      title={PAGURI: a user experience study of creative interaction with text-to-music models}, 
      author={Francesca Ronchini and Luca Comanducci and Gabriele Perego and Fabio Antonacci},
      year={2024},
      eprint={2407.04333},
      archivePrefix={arXiv},
      primaryClass={cs.SD},
      url={https://arxiv.org/abs/2407.04333}, 
}

@article{Likert1932,
  author  = {Likert, Rensis},
  title   = {A Technique for the Measurement of Attitudes},
  journal = {Archives of Psychology},
  number  = {140},
  pages   = {1--55},
  year    = {1932}
}

@article{Carifio2008Likert,
  author  = {Carifio, James and Perla, Rocco J.},
  title   = {Resolving the 50-year Debate Around Using and Misusing Likert Scales},
  journal = {Medical Education},
  volume  = {42},
  number  = {12},
  pages   = {1150--1152},
  year    = {2008}
}

@misc{Magenta2016,
  title        = {Magenta: Music and Art Generation with Machine Intelligence},
  author       = {{Google Brain Team}},
  howpublished = {\url{https://magenta.tensorflow.org/}},
  year         = {2016},
}

@article{jukebox2020,
  title   = {Jukebox: A Generative Model for Music},
  author  = {Dhariwal, Prafulla and Jun, Heewoo and Payne, Christine and Kim, Jong Wook and Radford, Alec and Sutskever, Ilya and Knight, Jeff},
  journal = {arXiv:2005.00341},
  year    = {2020},
  url     = {https://arxiv.org/abs/2005.00341}
}

@inproceedings{10.1145/3689050.3705995,
author = {Chang, Ethan and Chen, Zhixing and Labrune, Jb and Coelho, Marcelo},
title = {Be the Beat: AI-Powered Boombox for Music Suggestion from Freestyle Dance},
year = {2025},
isbn = {9798400711978},
publisher = {Association for Computing Machinery},
address = {New York, NY, USA},
url = {https://doi.org/10.1145/3689050.3705995},
doi = {10.1145/3689050.3705995},
booktitle = {Proceedings of the Nineteenth International Conference on Tangible, Embedded, and Embodied Interaction},
articleno = {67},
numpages = {6},
location = {
},
series = {TEI '25}
}

@misc{TonnAPI,
  title        = {Tonn API Documentation},
  howpublished = {\url{https://tonn.roexaudio.com/}},
  author       = {{RoEx Audio}},
  year         = {2024}
}

@misc{PyDub,
  title        = {pydub: Manipulate audio with a simple {Python} API},
  author       = {James Robert (jiaaro)},
  howpublished = {\url{https://github.com/jiaaro/pydub}},
  year         = {2011}
}

@inproceedings{McPherson2015Bela,
  author       = {McPherson, Andrew P. and Zappi, Victor},
  title        = {An Environment for Submillisecond-Latency Audio and Sensor Processing on BeagleBone Black},
  booktitle    = {Audio Engineering Society Convention 138},
  year         = {2015},
  month        = may,
  address      = {Warsaw, Poland},
  organization = {Audio Engineering Society},
  note         = {Convention Paper, presented May 7--10, 2015},
  url          = {https://instrumentslab.org/data/andrew/mcpherson_aes2015.pdf}
}

\newpage

\section{Appendix}
\begin{figure}[H]
    \centering
    \includegraphics[width=0.75\linewidth]{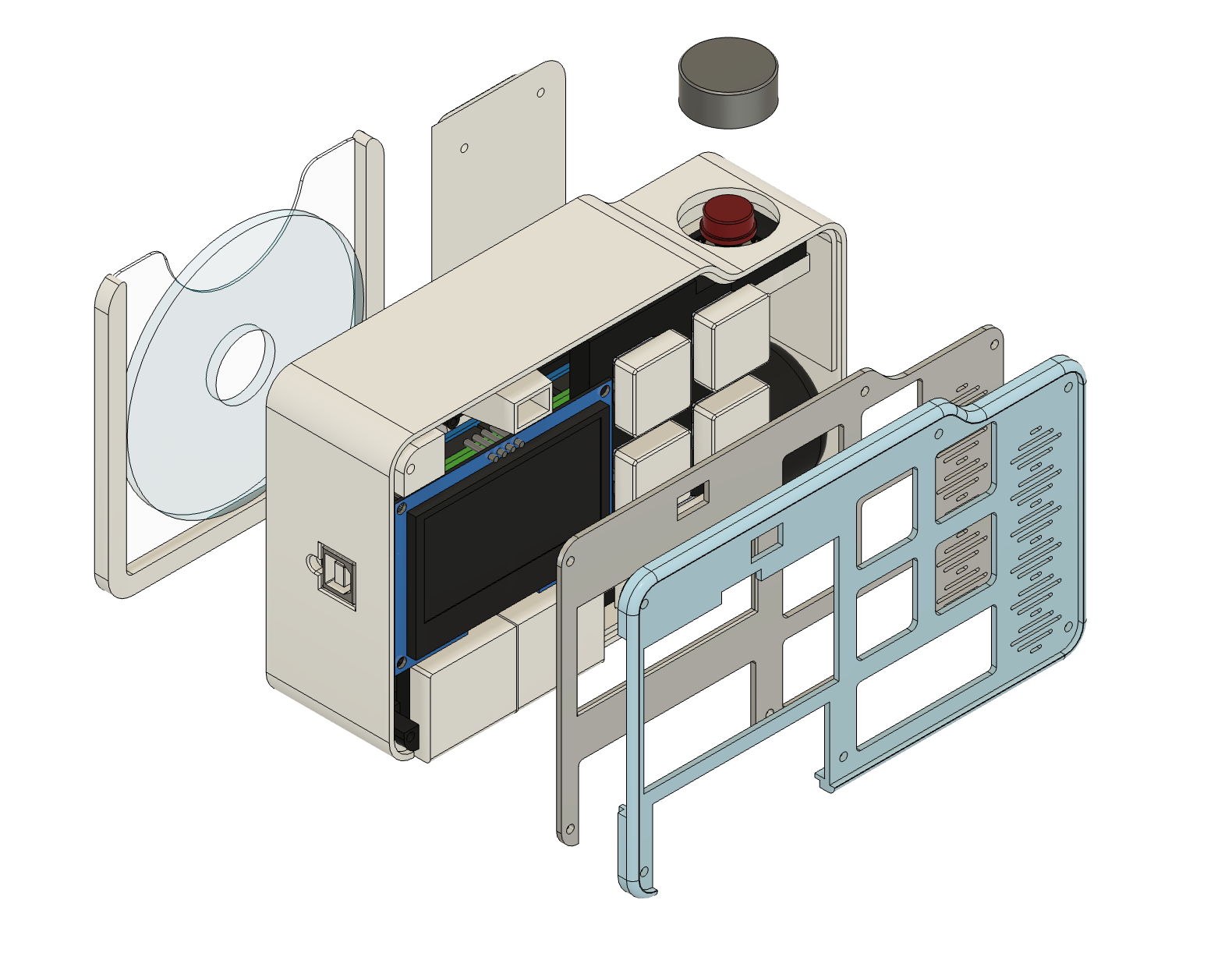}
    \caption{Interface exploded view}
    \label{fig:placeholder}
\end{figure}

\begin{figure}[H]
    \centering
    \includegraphics[width=0.75\linewidth]{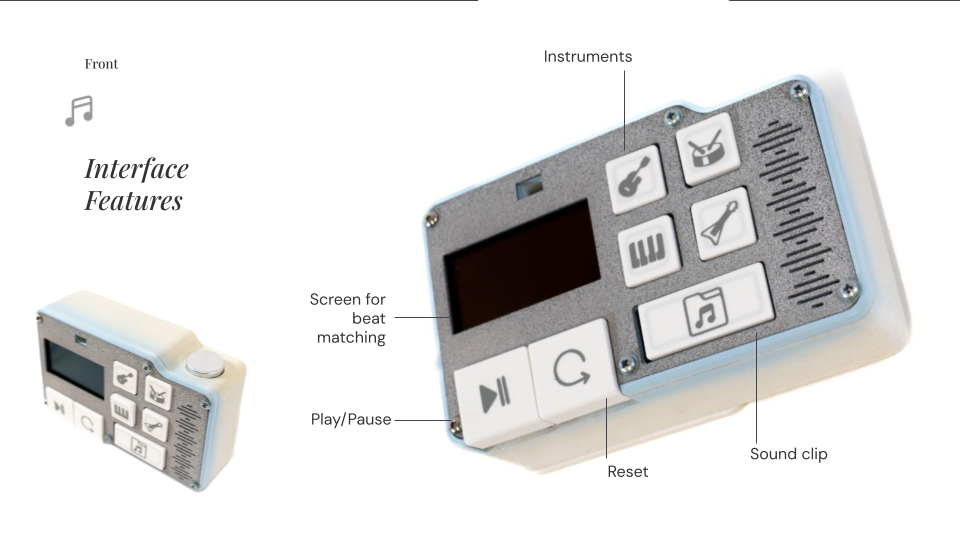}
    \caption{Front of LUMIA}
    \label{fig:placeholder}
\end{figure}

\begin{figure}[H]
    \centering
    \includegraphics[width=0.75\linewidth]{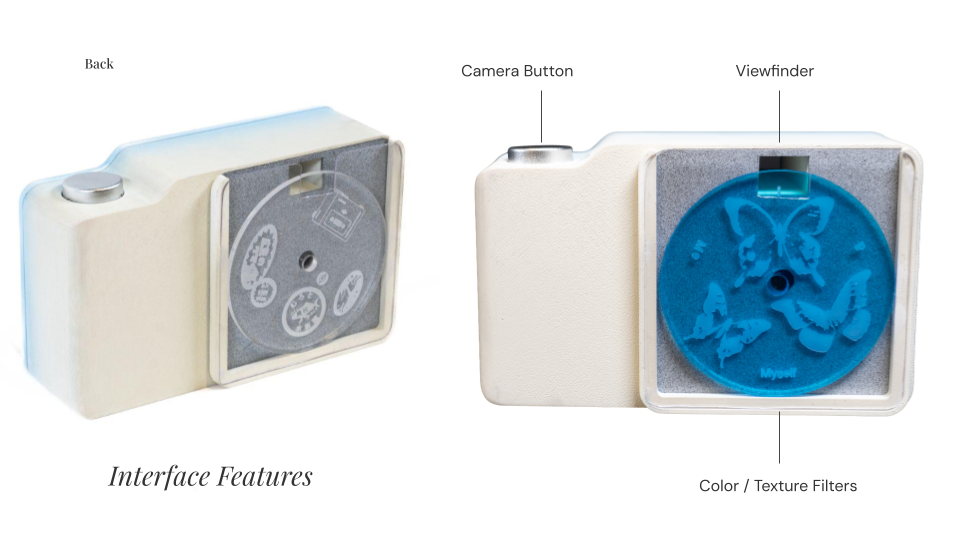}
    \caption{Back of LUMIA}
    \label{fig:placeholder}
\end{figure}

\begin{figure}[H]
    \centering
    \includegraphics[width=0.75\linewidth]{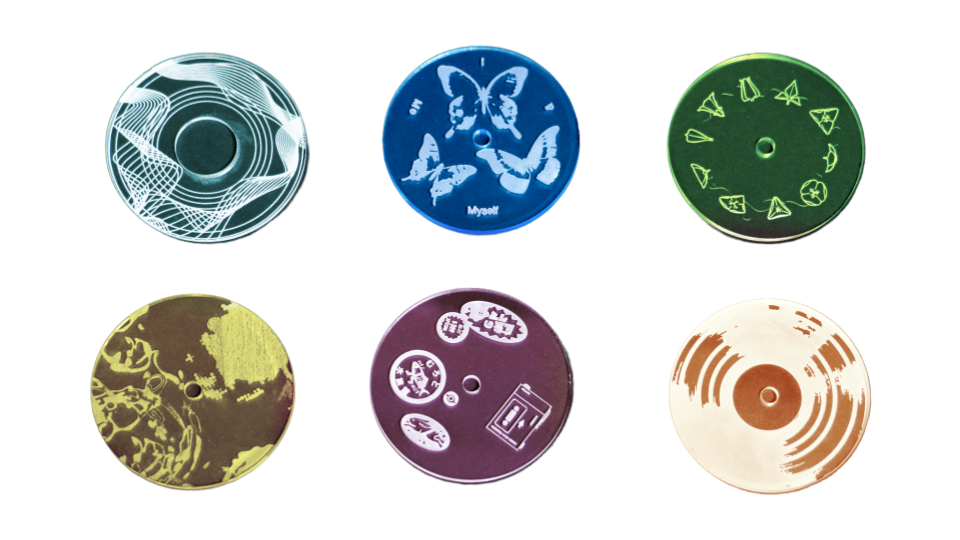}
    \caption{Prototyped Color Filter Disk Designs}
    \label{fig:placeholder}
\end{figure}

\begin{figure}[H]
    \centering
    \includegraphics[width=0.75\linewidth]{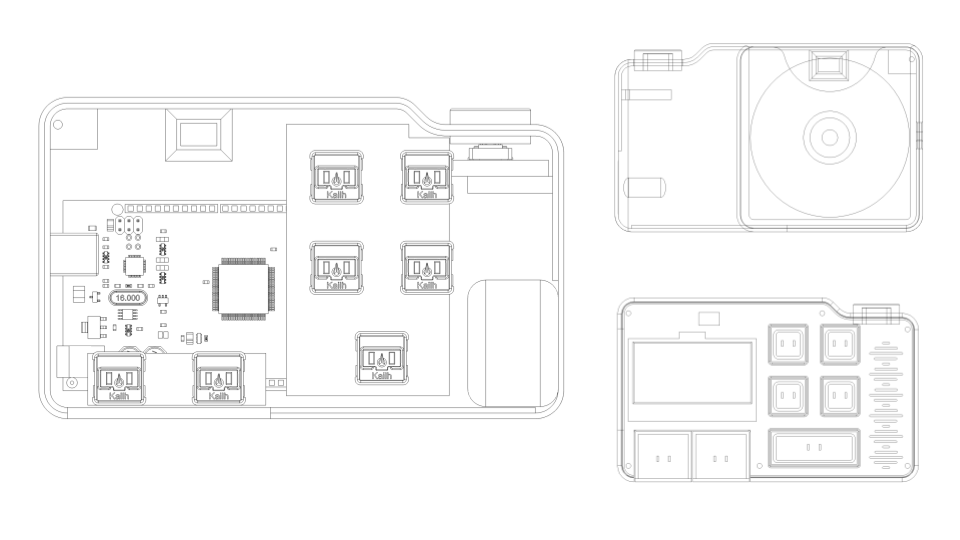}
    \caption{Schematics and Component Layout}
    \label{fig:placeholder}
\end{figure}

\begin{figure}[H]
    \centering
    \includegraphics[width=0.75\linewidth]{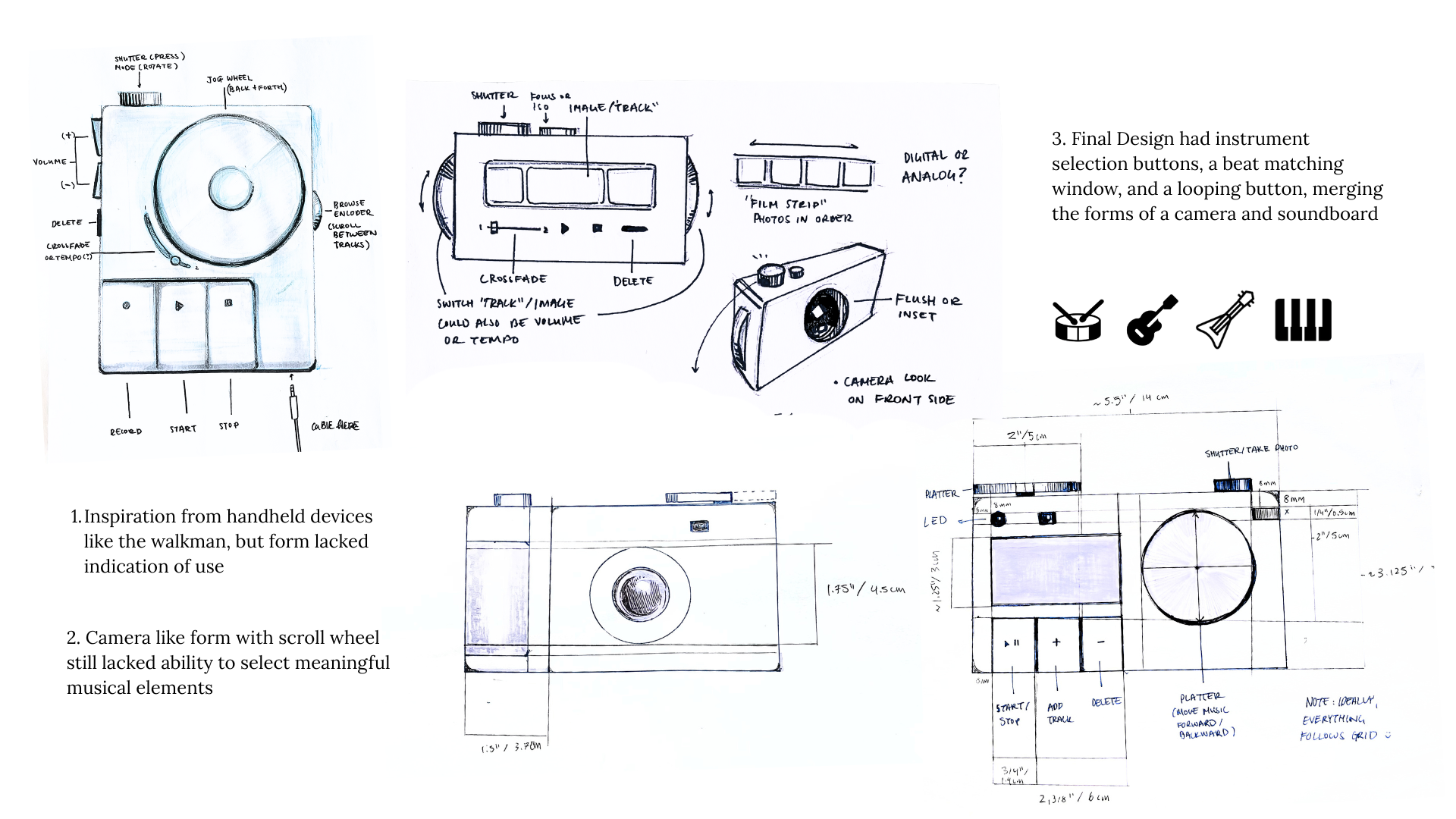}
    \caption{Form design evolution, driven by functionality and useability}
    \label{fig:placeholder}
\end{figure}

\subsection{Compute Resources and Latency}
\label{sec:compute}

\textit{Lumia}'s music generation pipeline integrates cloud-based multimodal AI services accessed via REST APIs, orchestrated by a local browser-based frontend. All experiments were conducted using the following configuration:

\paragraph{Hardware:}
\begin{itemize}
    \item \textbf{Frontend:} MacBook Pro (Apple M2 Pro, 16\,GB RAM) running Chrome~126
    \item \textbf{Microcontroller:} Arduino Nano 33 IoT (Cortex-M0+, 256\,KB SRAM) handling physical I/O
    \item \textbf{Network:} 1\,Gbps wired Ethernet or 300\,Mbps Wi-Fi~6 connection
\end{itemize}

\paragraph{Cloud Services:}
\begin{itemize}
    \item \textbf{Image captioning} --- OpenAI GPT-4 Vision API  
    Input: 1280$\times$720\,px JPEG frame ($\sim$120\,KB)  
    Average latency: $1.2 \pm 0.3$\,s  
    Cost: $\sim\$0.002$ per request ($\approx$120 input tokens, 200 output tokens)
    
    \item \textbf{Music generation} --- Stability AI Stable Audio 2.0 API  
    Input: single-sentence structured prompt ($\sim$35 tokens)  
    Output: 15\,s stereo WAV at 44.1\,kHz  
    Average latency: $3.8 \pm 0.6$\,s  
    Cost: $\sim\$0.14$ per generation (1 API credit)
    
    \item \textbf{Automated mixing \& mastering} --- Tonn Audio Mixing API  
    Input: up to 4 stereo stems, 15\,s each  
    Average latency: $5.2 \pm 0.9$\,s (mixpreview), $8.6 \pm 1.1$\,s (full master)  
    Cost: $\sim\$0.05$ per stem for mixpreview; $\sim\$0.15$ for final master
\end{itemize}

\paragraph{Execution time:}
\begin{itemize}
    \item End-to-end latency (capture $\rightarrow$ audio in loop): 5.0--6.5\,s
    \item End-to-end latency (capture $\rightarrow$ mixed audio update): 10--13\,s
\end{itemize}

\paragraph{Compute load:}
\begin{itemize}
    \item All ML inference performed on provider-side infrastructure (no GPU required locally)
    \item Frontend CPU load $\leq$ 8\% during playback; memory footprint $\sim$400\,MB with 4 concurrent AudioBuffers loaded
\end{itemize}

This configuration ensures reproducibility of performance results and can be replicated with comparable local hardware and a standard API subscription for the listed services.

\begin{figure}[H]
    \centering
    \includegraphics[width=0.75\linewidth]{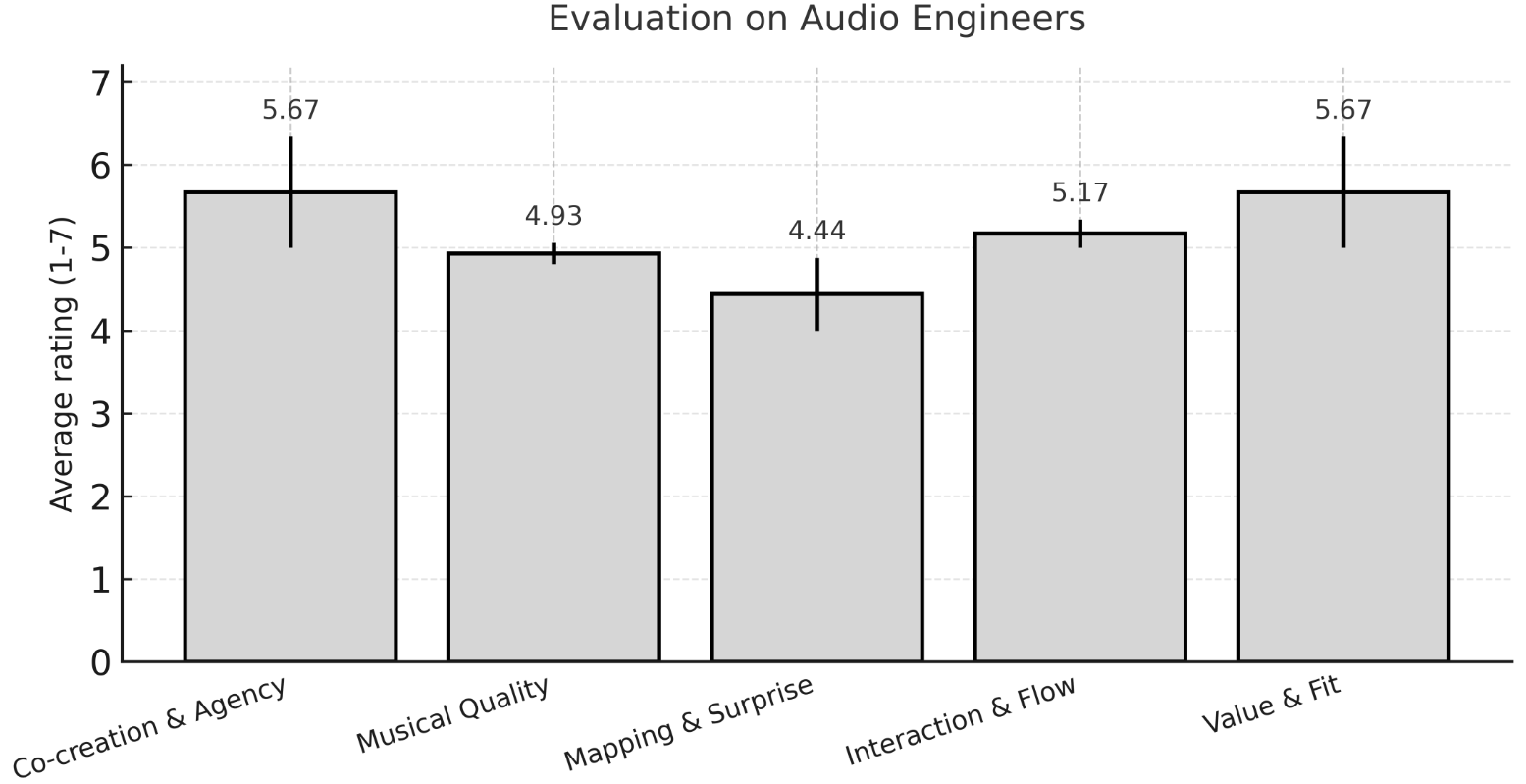}
    \caption{Collected numerical evaluation results}
    \label{fig:evaluation}
\end{figure}

\medskip

\newpage
\section*{NeurIPS Paper Checklist}
\begin{enumerate}

\item {\bf Claims}
    \item[] Question: Do the main claims made in the abstract and introduction accurately reflect the paper's contributions and scope?
    \item[] Answer: \answerYes{} 
    \item[] Justification:The abstract and introduction clearly state the system’s technical contributions, including the real-time vision-to-audio pipeline, the handheld interface design, and the focus on human–AI co-creation. These claims align directly with the methods, experiments, and results presented in the paper, without overstating scope or novelty beyond what is substantiated in the body.

\item {\bf Limitations}
    \item[] Question: Does the paper discuss the limitations of the work performed by the authors?
    \item[] Answer: \answerYes{} 
    \item[] Justification: The abstract and introduction accurately describe Lumia’s contributions, a real-time photo-to-music generation interface integrating GPT-4 Vision and Stable Audio, embedded in an embodied, camera-like device. The claims match the described methods, scope, and results without overstatement (Sections 1, 3–5).
    
\item {\bf Theory assumptions and proofs}
    \item[] Question: For each theoretical result, does the paper provide the full set of assumptions and a complete (and correct) proof?
    \item[] Answer:  \answerNA{} 
  
    \item {\bf Experimental result reproducibility}
    \item[] Question: Does the paper fully disclose all the information needed to reproduce the main experimental results of the paper to the extent that it affects the main claims and/or conclusions of the paper (regardless of whether the code and data are provided or not)?
    \item[] Answer: \answerYes{} 
    \item[] Justification: The paper provides sufficient methodological and implementation details to reproduce the pipeline, including hardware configuration (Section 3.1–3.2), system architecture (Section 4), vision-to-music pipeline parameters, audio generation settings, and evaluation design (Section 6). A link to the github repository containing written code is also included in the appendix, which contains more detailed instructions on how to run the pipeline.

\item {\bf Open access to data and code}
    \item[] Question: Does the paper provide open access to the data and code, with sufficient instructions to faithfully reproduce the main experimental results, as described in supplemental material?
    \item[] Answer: \answerYes{} 
    \item[] Justification: Code, data, and instructions are publicly available via the linked GitHub repository, system architecture and other details are described in the main paper.
 
\item {\bf Experimental setting/details}
    \item[] Question: Does the paper specify all the training and test details (e.g., data splits, hyperparameters, how they were chosen, type of optimizer, etc.) necessary to understand the results?
    \item[] Answer: \answerNA{} 

\item {\bf Experiment statistical significance}
    \item[] Question: Does the paper report error bars suitably and correctly defined or other appropriate information about the statistical significance of the experiments?
    \item[] Answer: \answerNA{} 
 
\item {\bf Experiments compute resources}
    \item[] Question: For each experiment, does the paper provide sufficient information on the computer resources (type of compute workers, memory, time of execution) needed to reproduce the experiments?
    \item[] Answer: \answerYes{} 
    \item[] Justification: All compute requirements are documented in Appendix, including local hardware (Arduino, browser frontend), cloud APIs used (GPT-4V, Stable Audio 2.0, Tonn), average API latency, and per-generation credit cost. No high-performance compute required beyond a standard laptop.
    
\item {\bf Code of ethics}
    \item[] Question: Does the research conducted in the paper conform, in every respect, with the NeurIPS Code of Ethics \url{https://neurips.cc/public/EthicsGuidelines}?
    \item[] Answer: \answerYes{} 
    \item[] Justification: The research complies with the NeurIPS Code of Ethics. All participants in the user study were informed of the research goals, data usage, and their right to withdraw; no personally identifiable information was collected or stored. The system poses no foreseeable harm, avoids biased or discriminatory model outputs by using generic prompts and instrumentation, and all AI services are used within their terms of service.

\item {\bf Broader impacts}
    \item[] Question: Does the paper discuss both potential positive societal impacts and negative societal impacts of the work performed?
    \item[] Answer: \answerYes{} 
    \item[] Justification: The paper addresses positive societal impacts, such as lowering barriers to music creation, enabling non-experts to engage in composition, and supporting new modes of co-creative expression. Potential negative impacts are also acknowledged, including possible overreliance on AI-generated content, the risk of homogenization of musical style due to model biases, and dependency on third-party AI services. These considerations are discussed in the context of responsible deployment and future work.
    
\item {\bf Safeguards}
    \item[] Question: Does the paper describe safeguards that have been put in place for responsible release of data or models that have a high risk for misuse (e.g., pretrained language models, image generators, or scraped datasets)?
    \item[] Answer: \answerNA{} 

\item {\bf Licenses for existing assets}
    \item[] Question: Are the creators or original owners of assets (e.g., code, data, models), used in the paper, properly credited and are the license and terms of use explicitly mentioned and properly respected?
    \item[] Answer: \answerYes{} 
    \item[] Justification: All external models, APIs, and datasets used in Lumia (GPT-4V, Stability AI’s Stable Audio, and Tonn mixing API) are credited in the paper with proper citations. Their use complies with the providers’ licensing and terms of service. No unlicensed or restricted assets were incorporated.
 
\item {\bf New assets}
    \item[] Question: Are new assets introduced in the paper well documented and is the documentation provided alongside the assets?
    \item[] Answer: \answerNA{} 

\item {\bf Crowdsourcing and research with human subjects}
    \item[] Question: For crowdsourcing experiments and research with human subjects, does the paper include the full text of instructions given to participants and screenshots, if applicable, as well as details about compensation (if any)? 
    \item[] Answer: \answerNA{} 

\item {\bf Institutional review board (IRB) approvals or equivalent for research with human subjects}
    \item[] Question: Does the paper describe potential risks incurred by study participants, whether such risks were disclosed to the subjects, and whether Institutional Review Board (IRB) approvals (or an equivalent approval/review based on the requirements of your country or institution) were obtained?
    \item[] Answer: \answerNA{} 

\item {\bf Declaration of LLM usage}
    \item[] Question: Does the paper describe the usage of LLMs if it is an important, original, or non-standard component of the core methods in this research? Note that if the LLM is used only for writing, editing, or formatting purposes and does not impact the core methodology, scientific rigorousness, or originality of the research, declaration is not required.
    \item[] Answer: \answerYes{} 
    \item[] Justification: The paper explicitly details the use of GPT-4 Vision, a large multimodal language model, as a core component of the vision-to-music pipeline. Its role in structured scene analysis, prompt construction, and genre/mood inference is described in the Methodology section, with technical specifications and integration details provided.

\end{enumerate}

\end{document}